\documentclass{article}

\usepackage[sglblindworkshop, final]{ai4nextg_neurips_2025}

\usepackage[utf8]{inputenc} % allow utf-8 input
\usepackage[T1]{fontenc}    % use 8-bit T1 fonts
\usepackage{hyperref}       % hyperlinks
\usepackage{url}            % simple URL typesetting
\usepackage{booktabs}       % professional-quality tables
\usepackage{amsfonts}       % blackboard math symbols
\usepackage{nicefrac}       % compact symbols for 1/2, etc.
\usepackage{microtype}      % microtypography
\usepackage{xcolor}         % colors
\usepackage{amsmath}
\usepackage{graphicx}
\usepackage{caption} % pour \captionof
\usepackage{float}
\usepackage{amssymb}
\usepackage[most]{tcolorbox}
\usepackage{enumitem}
\usepackage{subcaption}
\usepackage{parskip}
\usepackage{pifont}

\usepackage{tikz}
\usetikzlibrary{positioning}

\definecolor{viridis1}{HTML}{440154} % dark purple
\definecolor{viridis2}{HTML}{3B528B} % indigo
\definecolor{viridis3}{HTML}{21918C} % teal
\definecolor{viridis4}{HTML}{5EC962} % light green
\definecolor{viridis5}{HTML}{FDE725} % yellow (usually too light for white text)
\newcommand{\good}{\textcolor{viridis4!70!black}{\ding{51}}} % ✓
\newcommand{\bad}{\textcolor{red!70!black}{\ding{55}}}       % ✗
\newcommand{\Good}{\Large\good}
\newcommand{\Bad}{\Large\bad}

\tcbset{
  mybox/.style 2 args={
    enhanced,
    colback=#1!6,               % light background
    colframe=#1!75!black,
    boxrule=0.7pt,
    arc=10pt,
    left=6mm,right=6mm,top=2mm,bottom=2mm,
    drop shadow=black!20,
    attach boxed title to top left={xshift=10mm, yshift*=-\tcboxedtitleheight/2},
    boxed title style={
      colback=#1,
      colframe=#1!80!black,
      top=1pt,bottom=1pt,left=6pt,right=6pt,
      fontupper=\bfseries\color{#2}
    },
    before skip=8pt, after skip=8pt
  }
}

\newtcolorbox{qbox}{
  mybox={viridis2}{white},      % Question
  title={Prompt}
}
\newtcolorbox{vanillabox}{
  mybox={viridis3}{white},      % LLM (vanilla)
  title={GPT-4o (vanilla)}
}
\newtcolorbox{ragbox}{
  mybox={viridis4}{black},      % LLM + RAG (light background -> dark text)
  title={GPT-4o with RAG context}
}

\title{Retrieval-Augmented Generation for Reliable Interpretation of Radio Regulations}

\author{%
  Zakaria El Kassimi \\
  KAUST\\
  \texttt{zakaria.kassimi@kaust.edu.sa} \\
  \And
  Fares Fourati\\
  KAUST \\
  \texttt{fares.fourati@kaust.edu.sa} \\
  \AND
  Mohamed-Slim Alouini \\
  KAUST \\
  \texttt{slim.alouini@kaust.edu.sa} \\
}

\begin{document}

\maketitle

\begin{abstract}
We study question answering in the domain of radio regulations, a legally sensitive and high-stakes area. We propose a telecom-specific Retrieval-Augmented Generation (RAG) pipeline and introduce, to our knowledge, the first multiple-choice evaluation set for this domain, constructed from authoritative sources using automated filtering and human validation. To assess retrieval quality, we define a domain-specific retrieval metric, under which our retriever achieves approximately 97\% accuracy. Beyond retrieval, our approach consistently improves generation accuracy across all tested models. In particular, while naively inserting documents without structured retrieval yields only marginal gains for GPT-4o (less than 1\%), applying our pipeline results in nearly a 12\% relative improvement. These findings demonstrate that carefully targeted grounding provides a simple yet strong baseline and an effective domain-specific solution for regulatory question answering. All code and evaluation scripts, along with our derived question–answer dataset, are available at \url{https://github.com/Zakaria010/Radio-RAG}.
\end{abstract}

\section{Introduction}
Large Language Models (LLMs) have transformed natural language processing, achieving state-of-the-art performance in summarization, translation, and question answering. However, despite their versatility, LLMs are prone to generating false or misleading content, a phenomenon commonly referred to as \emph{hallucination} \cite{fourati2025coherence, Huang_2025, sahoo2024comprehensivesurveyhallucinationlarge}. While sometimes harmless in casual applications, such inaccuracies pose significant risks in domains that demand strict factual correctness, including medicine, law, and telecommunications. In these settings, misinformation can have severe consequences, ranging from financial losses to safety hazards and legal disputes.

\begin{figure*}[t]
    \centering
    \includegraphics[width=0.8\linewidth]{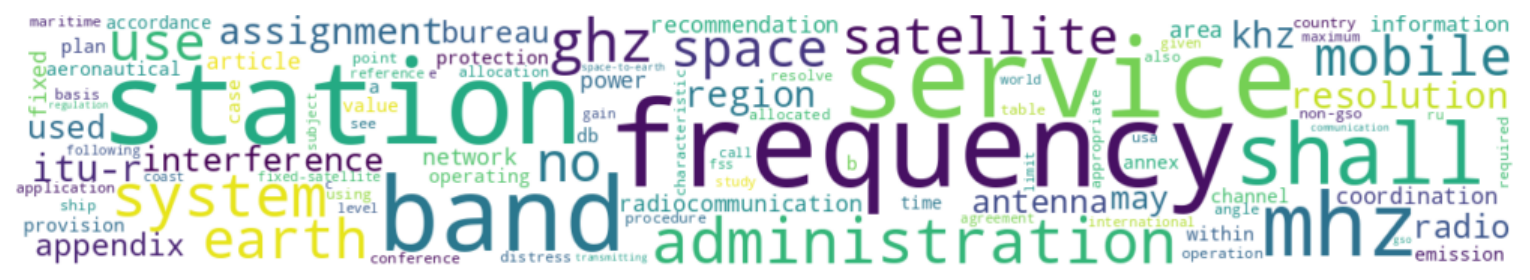}
    \caption{Distribution (word cloud) of key terms in the Radio Regulations corpus, highlighting the domain-specific vocabulary our pipeline must handle}
    \label{fig:right}
\end{figure*}

The telecommunications domain presents a particularly challenging case. Regulatory frameworks, and especially the ITU Radio Regulations \cite{ITURadioRegulations2024online}, are legally binding, technically intricate, and demand precise interpretation to ensure compliance. Even small errors can trigger costly service outages, legal disputes, or disruptions to critical infrastructure. Consequently, operators, regulators, and domain experts require reliable tools to assist in interpreting these regulations. As illustrated in Fig.~\ref{fig:right}, the Radio Regulations corpus exhibits a dense domain-specific vocabulary, underscoring why general-purpose LLMs often struggle in this setting and motivating the need for tailored approaches.

To address these challenges, we introduce a domain-specialized Retrieval-Augmented Generation (RAG) pipeline for regulatory question answering. Our system leverages authoritative external resources to ground LLM outputs, thereby reducing hallucinations and improving reliability \cite{lewis2021retrieval, gupta2024comprehensive}. To evaluate this approach, we construct a dedicated dataset of regulatory questions and answers derived from the ITU Radio Regulations \cite{ITURadioRegulations2024online}, validated through both automated checks and human review. To make this system directly accessible to practitioners, we deploy it as a custom \emph{Generative Pre-trained Transformer} (GPT) assistant, which integrates our RAG backend within a conversational interface.

Our contributions are fourfold:
\begin{enumerate}
    \item We design a RAG pipeline tailored for interpreting and answering regulatory inquiries in the telecommunications domain, with a focus on radio regulations.
    \item We create and release a curated dataset of question--answer pairs directly derived from the ITU Radio Regulations, rigorously validated for accuracy and completeness.
    \item We present an extensive empirical evaluation across multiple LLMs, showing that our pipeline consistently improves accuracy (e.g., +11.9\% absolute for GPT-4o \cite{hurst2024gpt}) and achieves up to 97\% retrieval accuracy.
    \item We deploy \textbf{Radio Regulations GPT}, a custom GPT front-end powered by our RAG backend, providing an interactive, reference-backed assistant for querying the ITU Radio Regulations.\footnote{Link to Radio Regulations GPT: \\ \url{https://chatgpt.com/g/g-687bac267f508191a97daf466eccfa50-radio-regulations-gpt}} 
\end{enumerate}

By enhancing the precision of automated regulatory interpretation, our work supports more reliable compliance, greater operational efficiency, and improved decision-making across the telecommunications sector.

\section{Background}

\subsection{LLMs}

LLMs are transformer-based neural networks \cite{vaswani2023attentionneed,xiao2025foundationslargelanguagemodels} trained on massive text corpora to acquire broad language understanding and reasoning capabilities. Recent models such as GPT-4 \cite{hurst2024gpt} and GPT-5 demonstrate remarkable progress, yet still exhibit hallucinations and limited long-term memory \cite{fourati2025coherence}. 

LLMs performance can be further enhanced through several complementary strategies. Supervised fine-tuning (SFT) adapts pretrained models to specific domains or tasks by updating their parameters on curated labeled data \cite{raffel2023exploringlimitstransferlearning,hu2021loralowrankadaptationlarge,xu2023parameterefficientfinetuningmethodspretrained}, while reinforcement learning from human feedback (RLHF) aligns their behavior with human preferences through reward-based optimization \cite{ouyang2022traininglanguagemodelsfollow}. Both methods directly modify the model’s parametric memory, embedding new knowledge into its weights, yet this knowledge remains static and can quickly become obsolete, especially in rapidly evolving or highly regulated domains \cite{ovadia2024finetuningretrievalcomparingknowledge}.

In contrast, prompt learning \cite{zhou2022large, kharrat2024acing} and prompt engineering \cite{sahoo2024systematic} guide model behavior without altering its parameters, using carefully crafted instructions or few-shot exemplars to induce desired task behavior at inference time. Relatedly, chain-of-thought (CoT) \cite{wei2022chain} prompting encourages models to generate intermediate reasoning steps before producing an answer, which improves multi-step reasoning but does not expand the model’s knowledge base. While these approaches enhance task adaptability and reasoning ability, they cannot update or extend the underlying knowledge stored in the model’s parameters, leaving LLMs vulnerable to hallucinations \cite{fourati2025coherence}, especially in specialized or fast-changing domains.

These limitations motivate the use of RAG, which complements LLMs and the above approaches with non-parametric memory by retrieving relevant external documents at inference time, thereby grounding their outputs in accurate, verifiable, and up-to-date information.

\subsection{RAG in General}

Retrieval-augmented generation (RAG) has emerged as a central paradigm for enhancing LLMs by conditioning them on external knowledge sources rather than relying solely on parametric memory. By retrieving relevant information from corpora, databases, or the web, RAG mitigates key limitations of static LLMs, including hallucination, factual drift, and domain obsolescence.

Recent surveys have mapped the evolution of RAG systems. Gao et al.~\cite{gao2024retrieval} and Wu et al.~\cite{wu2025retrieval} categorize approaches, analyzing design dimensions such as retrieval granularity (passages vs. documents), retriever–generator integration (late fusion vs. joint training), and memory augmentation. They also highlight persistent challenges, including reducing hallucinations, handling outdated or incomplete knowledge, and supporting efficient updates in dynamic domains. In parallel, Yu et al.~\cite{yu2024evaluation} emphasize the need for principled evaluation that captures the hybrid nature of retrieval and generation. Together, these works establish RAG as a promising direction for improving factuality, adaptability, and transparency in LLMs while underscoring unresolved research questions.

\subsection{Evaluating RAG}

Evaluating RAG remains a challenging open problem due to its hybrid nature. Yu et al.~\cite{yu2024evaluation} provide a comprehensive survey and propose \emph{Auepora}, a unified evaluation framework that organizes assessment along three axes: \textbf{Target}, \textbf{Dataset}, and \textbf{Metric}.  

On the \textit{Target} axis, they distinguish between component-level and end-to-end goals. For retrieval, the focus is on \emph{relevance} (alignment between query and retrieved documents) and \emph{accuracy} (alignment between retrieved content and ground-truth candidates). For generation, they emphasize \emph{relevance} (response–query alignment), \emph{faithfulness} (consistency with supporting documents), and \emph{correctness} (agreement with reference answers). Beyond these, they highlight broader system requirements such as low latency, diversity, robustness to noisy or counterfactual evidence, and calibrated refusal, which are critical for making RAG systems usable and trustworthy in practice.  

On the \textit{Dataset} axis, they note that widely used Q\&A-style benchmarks often capture only narrow aspects of RAG behavior, motivating domain-specific testbeds that reflect dynamic knowledge and application-specific constraints (e.g., legal, medical, or financial domains).  

On the \textit{Metric} axis, they collate traditional rank-based retrieval measures (e.g., MAP, MRR~\cite{tang2024multihoprag}), text generation metrics (e.g., ROUGE~\cite{ganesan2018rouge20updatedimproved}, BLEU~\cite{Papineni2002BleuAM}, BERTScore~\cite{zhang2020bertscoreevaluatingtextgeneration}), and the growing use of \emph{LLM-as-a-judge} for assessing faithfulness and overall quality. While LLM-based evaluation shows promise, they caution that alignment with human preferences and the need for transparent grading rubrics remain unresolved challenges.  

\noindent\textbf{Our evaluation choice.} In contrast to prior work that primarily reuses general-purpose benchmarks, we construct our own domain-specific testbed: a multiple-choice dataset derived directly from the ITU Radio Regulations \cite{ITURadioRegulations2024online} using an automated pipeline with LLM generation, automated judging, and human verification (Section~\ref{ss:dataset}). This design ensures that the benchmark reflects realistic regulatory queries while retaining ground-truth answers. Accordingly, our end-to-end metric is simply \emph{answer accuracy}. To disentangle retrieval from generation, we further introduce a domain-tailored retrieval metric aligned with component-level \emph{relevance} and \emph{accuracy}. Technical details of this metric are developed in Section~\ref{ss:retrieval}.

\section{Related Works}
\begin{figure*}[t]
    \centering
    \includegraphics[width=\linewidth]{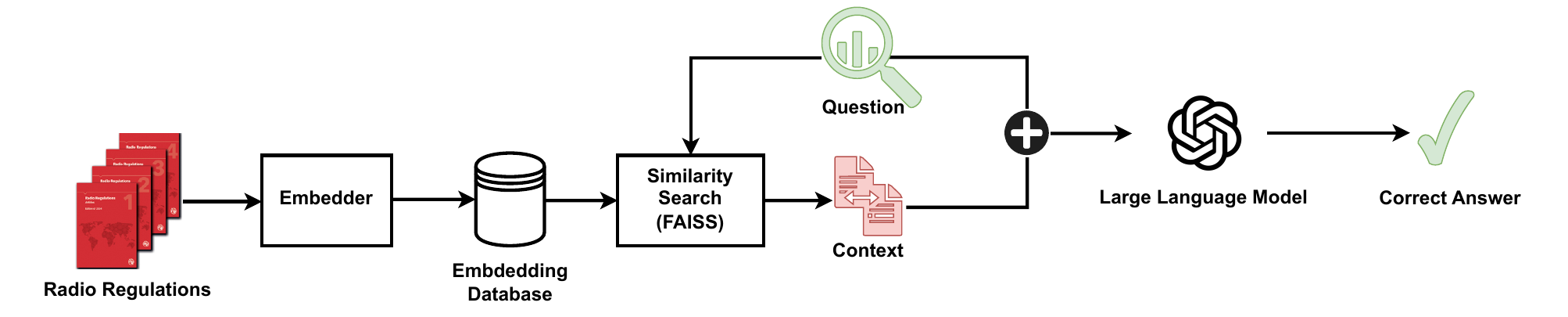}
    \caption{Overview of our Retrieval-Augmented Generation (RAG) pipeline for radio regulations QA, combining FAISS-based retrieval with LLM-based answer generation}
    \label{fig:left}
\end{figure*}

The telecommunications domain poses distinctive challenges for language models due to its dense technical standards, highly structured documents, and precise terminology requirements. Recent efforts have begun adapting Retrieval-Augmented Generation (RAG) frameworks to this space. Bornea et al.~\cite{bornea2024telcorag} introduce \emph{Telco-RAG}, a system designed to process 3GPP specifications, while Saraiva et al.~\cite{saraiva2024telcodpr} propose \emph{Telco-DPR}, which evaluates retrieval models on a hybrid dataset combining textual and tabular inputs from telecom standards. Maatouk et al.~\cite{maatouk2023teleqna} contribute \emph{TeleQnA}, a benchmark for assessing LLM knowledge of telecommunications. Subsequently \emph{Tele-LLMs}~\cite{maatouk2025telellmsseriesspecializedlarge} were presented as a family of domain-specialized models trained on curated datasets (Tele-Data) and evaluated on a large-scale benchmark (Tele-Eval), demonstrating that targeted pretraining and parameter-efficient adaptation yield substantial improvements over general-purpose LLMs. More recently, Zou et al.~\cite{zou2024telecomgpt} developed TelecomGPT, a telecom‐specific LLM built via continual pretraining, instruction tuning, and alignment tuning, and evaluated on new benchmarks such as Telecom Math Modeling, Telecom Open QnA, and Telecom Code Tasks; TelecomGPT outperforms general-purpose models including GPT-4, Llama-3, and Mistral in several telecom‐domain tasks. 

Despite these advances, no publicly available benchmark directly addresses radio regulations, a legally binding and technically demanding domain. Existing resources focus on telecom standards, network operations, or numerical reasoning, but none capture spectrum compliance, licensing rules, interference constraints, or jurisdictional variation. This gap underscores the novelty of our contribution: we introduce the first dataset of multiple-choice questions (MCQ) derived from the ITU Radio Regulations \cite{ITURadioRegulations2024online}, constructed through automated generation, LLM-based judging, and human verification. The dataset not only enables systematic evaluation of RAG in this domain but also provides a reusable testbed for future research.  

In contrast to prior work such as Telco-RAG and Telco-DPR, which target 3GPP specifications and generic telecom retrieval, our work focuses explicitly on radio regulations. Furthermore, we show that layering our RAG pipeline onto Tele-LLMs yields accuracy gains without any additional pretraining or fine-tuning, highlighting that carefully designed retrieval and grounding, rather than scaling parametric knowledge, is the key driver of performance in this setting.

\color{black}

\section{Methodology} 

Our RAG pipeline, as shown in Fig.~\ref{fig:left}, comprises two important sequential blocks: \emph{retrieval} and \emph{generation}, to answer radio regulations questions by grounding the LLM in relevant corpus excerpts.

\subsection{Retrieval Block}
\label{ss:retrieval}
We compute dense sentence embeddings for each corpus segment using the Sentence-Transformers model all-MiniLM-L6-v2 \cite{reimers2019sentencebert, wang2020minilm}, then build a Facebook AI Similarity Search (FAISS) \cite{douze2025faisslibrary} index over these vectors. At inference, we retrieve the top-$k$ most similar segments to the user query, where $k$ is a tunable parameter. This ensures the generator is supplied only with the most pertinent clauses.

To disentangle retrieval quality from downstream generation, we evaluate the retriever against a ground-truth context. For each question $i$, let $R_i$ denote the concatenation of the $k$ retrieved chunks and $C_i$ the ground-truth supporting context from our dataset. We compute the ROUGE-L \cite{ganesan2018rouge20updatedimproved} F$1$ score $F{1}^{(i)}$ between $R_i$ and $C_i$. A retrieval is considered correct if
\begin{equation}
F_{1}^{(i)} \ge \alpha,
\quad
\alpha \triangleq \gamma,F_{1,\max},
\quad
F_{1,\max} \triangleq
\frac{2,\min(R,,C)}{R + C},
\end{equation}
where $R$ and $C$ are the lengths of $R_i$ and $C_i$, respectively. That is a retrieved context is accepted as correct if it achieves at least a ratio $\gamma$ of the maximum achievable ROUGE-L overlap with the ground truth, which reflects the practical requirement that an answer does not demand the entire reference passage, capturing a sufficiently overlapping subset is often enough. In particular, the documents contain redundant clauses, so a strict exact-match requirement would underestimate retrieval quality.

Finally, retrieval accuracy is computed as the fraction of correctly retrieved instances:
\begin{equation}
  \mathrm{Acc}_{\mathrm{ret}}
  \triangleq
  \frac{1}{N}
  \sum_{i=1}^{N}
    \mathbf{1}\bigl(F_{1}^{(i)} \ge \alpha\bigr).
\end{equation}

\begin{figure}
    \centering
    \includegraphics[width=\linewidth]{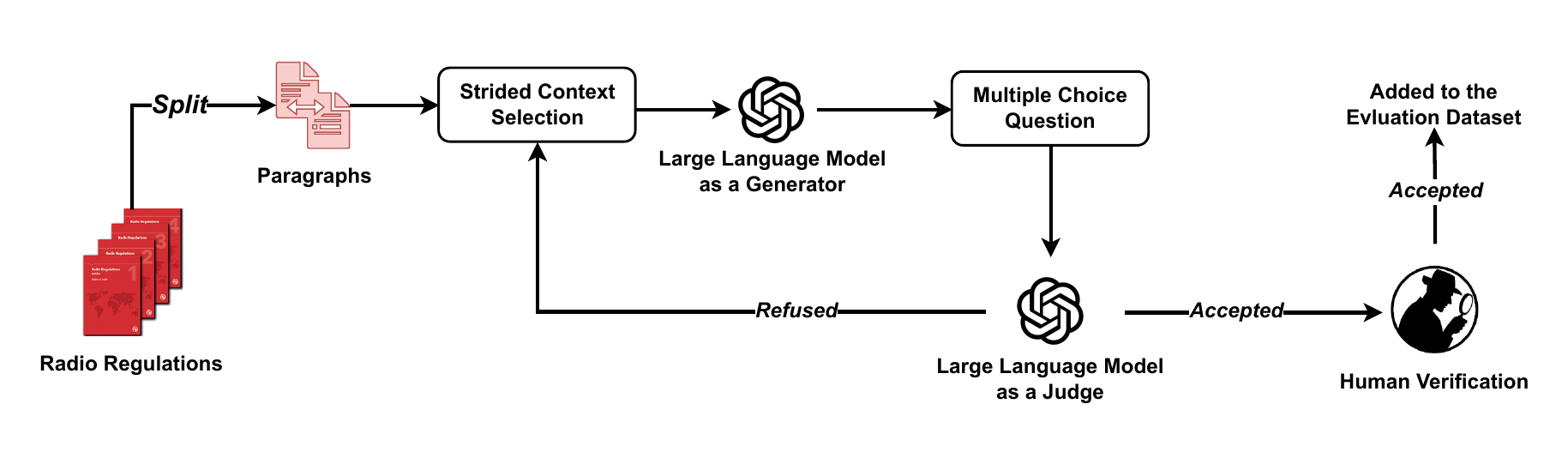}
    \caption{Automated pipeline for generating and validating multiple-choice questions from radio regulations, integrating LLM generation, automated judging, and human verification}
    \label{fig:eval_generations}
\end{figure}

\subsection{Reranking}
Before giving the retrieved chunks to the generator, we optionally use an LLM-based reranker. The goal of this reranker is to provide the generator with the best context in the best order depending on the given query. In our experiments provided in Table~\ref{tab:deepseekragrerank}, reranking yields a modest gain of about \(1\%\) absolute accuracy (at the cost of roughly \(1.5\times\) higher end-to-end compute time). We disable it by default and recommend enabling it only when compute is abundant and accuracy gains are critical.

\subsection{Generation Block}
The $k$ retrieved paragraphs plus the MCQ prompt are concatenated into
\[
  \underbrace{\texttt{Paragraph 1: }r_{i,1}\;\dots\;\texttt{Paragraph k: }r_{i,k}}_{\text{context}}
\]
and prefaced with a system instruction:
``You are a radio regulations expert. Answer using the context.''
We then generate the answer with the chosen model among several models.

\subsection{Evaluation}
Since the task is MCQ-based, end-to-end performance is measured by standard accuracy:
\begin{equation}
  \mathrm{Acc}_{\mathrm{MCQ}}
  \triangleq 
  \frac{1}{N}
  \sum_{i=1}^{N}
    \mathbf{1}\bigl(\hat{a}_i = a_i\bigr),
\end{equation}
where $\hat{a}_i$ is the model’s selected option and $a_i$ the ground-truth.

By combining these blocks, we can both maximize answer correctness and pinpoint whether any errors arise from retrieval or generation.

\subsection{Dataset Construction}
\label{ss:dataset} 

We developed a detailed set \footnote{More details about our dataset are available at \url{https://github.com/Zakaria010/Radio-RAG}.} of questions specifically targeting radio regulations. As summarized in Fig.~\ref{fig:eval_generations}, this process involved extracting clean text from official regulatory documents and automatically generating realistic, domain-relevant questions and answers.

\paragraph*{1) Text Extraction and Chunking}  
We extract the full text from telecom-regulation PDF references, then segment it into paragraphs. Let \(T\) be the total number of words and \(M\) the number of segments.

\paragraph*{2) Uniform Sampling of Chunks}  
To maximize coverage, we sample segments in a two-pass strided fashion:
\begin{itemize}
  \item \textbf{First pass:} indices \(0, s, 2s, \dots\) with stride \(s = \max\{1, \lfloor M/N \rfloor\}\), where \(N\) is the target question count.
  \item \textbf{Second pass:} for each offset \(o = 1, \ldots, s-1\), indices \(o,\, o+s,\, o+2s, \dots\) until \(N\) accepted questions are reached.
\end{itemize}

\paragraph*{3) Question Generation}  
Each chunk is provided to a text-to-text LLM \texttt{google/flan-t5-xxl} \cite{chung2022scalings} with a rigorously defined prompt template enforcing the format:
\[
\begin{aligned}
&\texttt{Q: \emph{<question>}?}\\
&\texttt{Options: A) ... \;|\; B) ... \;|\; C) ... \;|\; D) ...}\\
&\texttt{Answer: \emph{<correct option>}}\\
&\texttt{Explanation: \emph{<justification>}}
\end{aligned}
\]

\paragraph*{4) Quality Filtering}  
Generated Q\&A entries are evaluated by a domain-expert judge model \texttt{AliMaatouk/Llama-3-8B-Tele} \cite{maatouk2025telellmsseriesspecializedlarge}. Only entries judged ``Good'' are retained; others re-enter the sampling loop.

While this four-step pipeline provides broad coverage of the regulatory corpus, it does not yet include a formal expert-verification stage. We therefore conduct a light human pass to remove obviously illogical questions; incorporating systematic expert review would further improve precision and will become increasingly valuable as the dataset grows. 
\begin{figure*}[t]
  \centering
  \includegraphics[width=\linewidth]{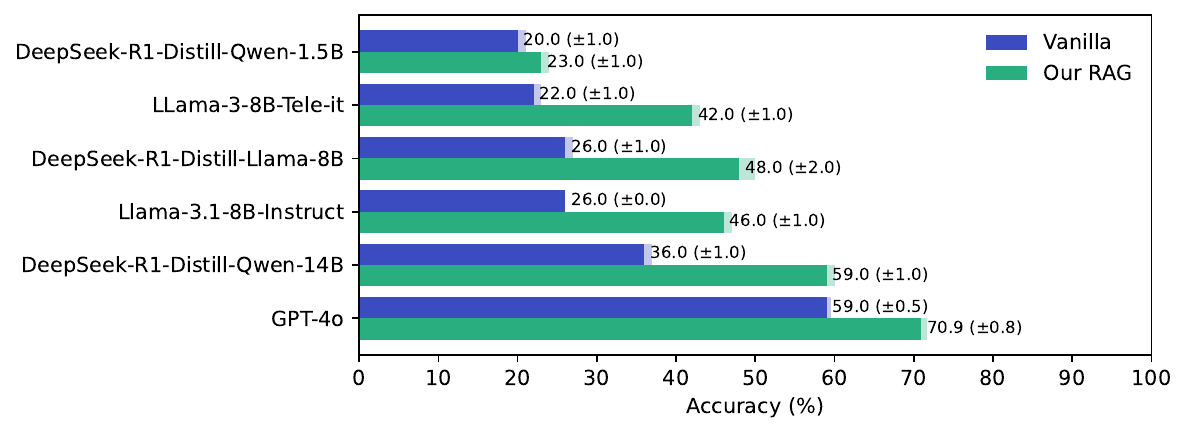}
  \caption{Accuracy comparison of vanilla LLMs versus our RAG-augmented approach, showing consistent gains across models, with GPT-4o achieving the largest improvement}
  \label{fig:comp}
\end{figure*}

\section{Results and Impact}
\subsection{Setup}

All experiments were run on a Slurm-managed HPC cluster using a single NVIDIA A100 GPU and 200\,GB of host memory per job. Retrieval used FAISS-GPU with sentence embeddings from the all-MiniLM-L6-v2 model, indexed over our corpus. Generation used open-source models locally and GPT-4o for comparison with one of the most popular models; the reranker was disabled by default and enabled only in the ablation. We report mean accuracy with standard error across runs, fixing random seeds for repeatability.

\subsection{The RAG Results}

Our experiments presented in Fig.~\ref{fig:comp} demonstrate that integrating RAG significantly improves model accuracy across all tested LLMs. Notably, GPT-4o~\cite{hurst2024gpt} exhibited a +11.9\% improvement, suggesting that even sophisticated commercial models significantly benefit from structured retrieval of regulatory contexts. DeepSeek-R1-Distill-Qwen-14B~\cite{guo2025deepseek} exhibited the largest absolute improvement, +23\%. Smaller models like DeepSeek-R1-Distill-Qwen-1.5B~\cite{guo2025deepseek} showed modest but notable gains of +3\%, indicating that retrieval contexts help models of all scales, albeit differently. Examples are provided in the Appendix, Section \ref{sec:ex}.

Interestingly, as shown in Table~\ref{tab:gpy_results}, directly uploading regulatory documents into the prompt, namely GPT-4o + full documents without retrieval, barely improved accuracy, underscoring that structured retrieval is crucial. 

\begin{table}[t]
    \centering
    \caption{Accuracy of GPT-4o with and without RAG augmentation (mean $\pm$ standard error across runs).}
    \begin{tabular}{lcc}
        \toprule
        Method                         & Accuracy         & Gain (\%) \\
        \midrule
        ChatGPT-4o                     & $59.0\% \pm 0.5$  & ---       \\
        ChatGPT-4o + full documents (no retrieval) & $59.6\% \pm 0.4$  & +0.6    \\
        ChatGPT-4o + our RAG           & $\textbf{70.9\%} \pm \textbf{0.8}$  & \textbf{+11.9}    \\
        \bottomrule
    \end{tabular}
    \label{tab:gpy_results}
\end{table}

\subsection{Retrieval-Only Results}

Table~\ref{tab:retrieval_results} summarizes results evaluating exclusively the retrieval component, isolating its performance from downstream generation tasks. The table reports retrieval accuracy across various configurations from a comprehensive hyperparameter sweep. Accuracy was measured using a ROUGE-L F$_1$ similarity threshold set at 0.7. We observe that configurations with chunk sizes of 500 and 700 words and higher values of top-$k$ consistently achieve superior retrieval performance, notably 97\% accuracy for the configuration with 700-word chunks and top-7 retrieval. Configurations with smaller chunks, 150 and 300 words, or fewer retrieved contexts, top-$k$ = 3, yielded significantly lower accuracy, demonstrating the importance of sufficient context for capturing regulatory information.

\section{Discussion}

A key strength of our approach is its robustness to regulatory updates and corpus expansion. Unlike supervised fine-tuning or RLHF, which embed knowledge directly into model parameters and require costly retraining when new information emerges, our RAG pipeline decouples knowledge from the model. The LLM remains fixed while the retriever dynamically accesses an external corpus that can be updated at any time. As new editions of the ITU Radio Regulations are released or additional documents become relevant, they can simply be indexed without retraining, preserving system validity as the regulatory landscape evolves. This modular design also scales naturally to larger corpora, since retrieval performance depends on embedding quality and index structure rather than model size.

Beyond its technical architecture, our work also contributes the first dedicated evaluation dataset for radio regulations, which can serve as a reusable benchmark for future research. By providing standardized, validated multiple-choice questions derived directly from authoritative sources, this dataset enables systematic comparison of retrieval and generation methods in a legally sensitive domain where no prior benchmark existed. It can support the development of domain-specific models, guide fine-tuning or retrieval design, and foster reproducibility by offering a stable testbed for future studies.

Finally, to illustrate the practicality of our approach, we deployed the pipeline as a custom GPT, named Radio Regulations GPT, integrated with ChatGPT via an Actions endpoint. This deployment demonstrates that our RAG architecture can serve as a reliable conversational assistant for practitioners, maintaining grounding in the official ITU corpus while remaining easily updatable as regulations evolve. For more details refer to Appendix~\ref{deployment}.

\section{Conclusion}

We presented a domain-specialized RAG framework and the first multiple-choice benchmark for interpreting radio regulations, showing that structured retrieval yields significant gains over vanilla prompting, while naive document insertion provides little benefit. By isolating retrieval from generation, we demonstrated that retrieval can be made reliably accurate. Our findings establish targeted grounding as a simple yet powerful baseline for legally sensitive domains and open directions for advancing generation strategies, refining reranking, and expanding human-verified regulatory datasets. Ultimately, this work underscores RAG's potential to make AI-assisted compliance both more accurate and more trustworthy in high-stakes telecom applications.

\newpage
\bibliographystyle{plain}
\bibliography{references}
\newpage
\appendix

\section{Additional Results}

\begin{table}[ht]
    \centering
    \caption{Impact of retrieval hyperparameters on RAG accuracy (DeepSeek-R1-Distill-Qwen-14B). Accuracy remains robust across FAISS indexing backends, but chunk size and top-k choices significantly affect performance, with insufficient retrieval (large chunks, low k) leading to accuracy drops.}
    \label{tab:abl_study}
    \begin{tabular}{@{}ccccc@{}}
        \toprule
        Index & Chunk Size & Overlap & Top-$k$ & Accuracy \\ \midrule
        flatl2 & 400 & 50 & 5 & \textbf{59\% ± 1} \\
        flatl2 & 1000 & 50 & 5 & 58\% ± 1 \\
        innerproduct & 400 & 50 & 3 & 57\% ± 1 \\
        innerproduct & 400 & 50 & 5 & 58\% ± 0 \\
        innerproduct & 700 & 50 & 3 & \textbf{59\% ± 1} \\
        innerproduct & 700 & 50 & 5 & 57\% ± 3 \\ 
        innerproduct & 1000 & 50 & 3 & 54 ± 0 \\ 
        innerproduct & 1000 & 50 & 5 & \textbf{59\% ± 1} \\ \bottomrule
    \end{tabular}
\end{table}

According to Table~\ref{tab:abl_study}, across indexing and context settings, accuracy remains tightly clustered around $57\%-59\%$, with tree configurations reaching $59\% \pm 1$. Switching the FAISS backend from inner product to flat L2 does not materially change outcomes given overlapping standard errors, indicating that the index choice is not the limiting factor. The only noticeable dip is at chunk size 1000 with top-k=3 ($54\% \pm 0$),suggesting insufficient evidence when k is too small for long chunks. With large chunks, increasing k to 5 recovers performance to $59\%$, while with medium chunks ($700$) a smaller k = 3 avoids a redundant or noisy context and performs best.

Overall, the ablation shows that our RAG is robust as long as the generator receives enough context.

\begin{table}[ht]
    \centering
    \caption{Accuracy of DeepSeek-R1-Distill-Qwen-14B with and without RAG/reranking (mean $\pm$ standard error across runs).}
    \label{tab:deepseekragrerank}
    \begin{tabular}{lcc}
        \toprule
        Method & Accuracy & Gain (\%) \\
        \midrule
        DeepSeek-R1-Distill-Qwen-14B (no RAG) & $36.0\% \pm 1.0$ & --- \\
        DeepSeek-R1-Distill-Qwen-14B + RAG (no reranking)   & $59.0\% \pm 1.0$ & +23.0 \\
        DeepSeek-R1-Distill-Qwen-14B + RAG (with reranking) & $\textbf{60.0\%} \pm \textbf{1.0}$ & \textbf{+24.0} \\
        \bottomrule
    \end{tabular}
\end{table}

\begin{table}[ht]
    \centering
    \caption{Retrieval-only accuracy across chunk sizes and top-k settings. Larger chunk sizes (500–700 characters) with higher top-k achieve >95\% accuracy, while small chunks (<300) fail to capture sufficient context, confirming the importance of retrieval granularity.}
    \label{tab:retrieval_results}
    \begin{tabular}{@{}cccc@{}}
        \toprule
        Chunk Size & Overlap & Top-$k$ & Accuracy \\ \midrule
        500 & 50 & 5 & 93\% \\
        500 & 50 & 7 & 95\% \\
        700 & 50 & 3 & 92\% \\
        700 & 50 & 5 & 95\% \\
        700 & 50 & 7 & \textbf{97}\% \\
        300 & 30 & 3 & 0\% \\
        300 & 30 & 5 & 73\% \\
        300 & 30 & 7 & 91\% \\
        150 & 10 & 3 & 0\% \\
        150 & 10 & 5 & 0\% \\
        150 & 10 & 7 & 1\% \\ \bottomrule
    \end{tabular}
\end{table}

\section{Deployment as Radio Regulations GPT}
\label{deployment}

% \subsection{Deployment as Radio Regulations GPT}
% \label{subsec:radio-reg-gpt}

To make our system easily accessible to practitioners and regulators, we deploy the proposed RAG pipeline as a custom GPT, denoted \emph{Radio Regulations GPT}. The RAG backend is exposed as a FastAPI service with a simple retrieval endpoint described via an OpenAPI schema:

\texttt{POST /query}: takes a JSON payload with the user question (\texttt{query}) and a fixed number of retrieved chunks (\texttt{top\_k}), and returns the most relevant snippets from the indexed ITU Radio Regulations.

This API is integrated into ChatGPT through the Actions interface, so that the Custom GPT can call our backend directly. We configure Radio Regulations GPT with explicit system instructions that:
(i) detect when a user request concerns the ITU Radio Regulations or related spectrum-policy questions;
(ii) in such cases, call the \texttt{/query} endpoint with the full user message as input;
(iii) treat the returned snippets as the primary evidence for the answer and synthesize a concise response grounded in these passages; 
(iv) include precise references to the relevant RR articles, numbers, and tables; and 
(v) fall back to standard LLM behavior without calling the API for clearly out-of-scope queries.
This deployment ensures that answers remain faithful to the official ITU text while providing a convenient conversational interface.

\section{Question-Answer Examples}
\label{sec:ex}
\begin{figure}[t]
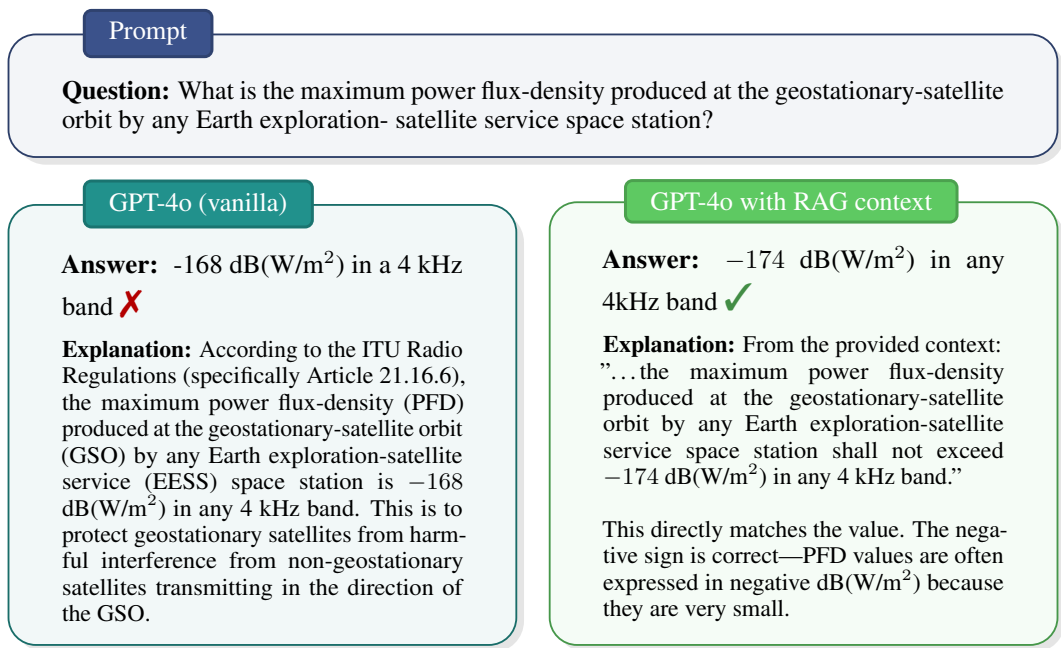

\centering

% --- Row 1: full-width question + options ---
\begin{subfigure}{\linewidth}
\begin{qbox}
\textbf{Question:} What is the maximum power flux-density produced at the
geostationary-satellite orbit by any Earth exploration- satellite service space station?
\end{qbox}
\end{subfigure}

\vspace{0.7em}

% --- Row 2: two columns for answers ---
\begin{subfigure}{0.485\linewidth}
\begin{vanillabox}
\textbf{Answer:} \ -168 dB(W/m$^2$) in a 4 kHz band  \Bad

\medskip
\small \textbf{Explanation:}
According to the ITU Radio Regulations (specifically Article 21.16.6), the maximum power flux-density (PFD) produced at the geostationary-satellite orbit (GSO) by any Earth exploration-satellite service (EESS) space station is $-168$ dB(W/m$^2$) in any 4 kHz band. This is to protect geostationary satellites from harmful interference from non-geostationary satellites transmitting in the direction of the GSO.
\end{vanillabox}
\end{subfigure}\hfill
\begin{subfigure}{0.485\linewidth}
\begin{ragbox}
\textbf{Answer:} $-174$ dB(W/m$^2$) in any 4kHz band  \Good

\medskip
\small \textbf{Explanation:} From the provided context:
\small 

''\ldots the maximum power flux-density produced at the geostationary-satellite orbit by any Earth exploration-satellite service space station shall not exceed $-174$ dB(W/m$^2$) in any 4 kHz band.''\\

This directly matches the value. The negative sign is correct—PFD values are often
expressed in negative dB(W/m$^2$) because they are very small.
\end{ragbox}
\end{subfigure}

\caption{Qualitative comparison of vanilla GPT-4o versus our RAG-augmented approach on a regulatory question, where RAG retrieves the rule and yields the correct answer.}
\label{fig:pfd-qa}
\end{figure}

The figure \ref{fig:pfd-qa} shows an example of how RAG efficiently helped GPT-4o \cite{hurst2024gpt} to get the correct answer.

Figures \ref{fig:bring-back-into-use} and \ref{fig:haps-band} show qualitative MCQ cases; bringing-into-use date for GSO assignments and the HAPS identification band; where our RAG-augmented GPT-4o retrieves the rule and answers correctly, unlike vanilla GPT-4o.

\begin{figure}[htbp]
\centering

% --- Row 1: full-width question + options ---
\begin{subfigure}{\linewidth}
\begin{qbox}
\textbf{Question:} What is the date of bringing back into use of the frequency assignment to a space station in the geostationary-satellite orbit?

\medskip
\textbf{Options}
\begin{enumerate}[label=\textbf{\Alph*.}, leftmargin=*, itemsep=2pt, topsep=2pt]
  \item The date on which the recorded assignment is brought back.
  \item The day on the commencement of 90-day period defined below.
  \item The end of 30 days from the end.
  \item The time on 15 days after the completion of.
\end{enumerate}
\end{qbox}
\end{subfigure}

\vspace{0.5em}

% --- Row 2: two columns for answers ---
\begin{subfigure}{0.485\linewidth}
\begin{vanillabox}
\textbf{Answer:} A.\ The date on which the recorded assignment is brought back. \Bad

\medskip
\small \textbf{Explanation:}
According to the ITU Radio Regulations, specifically the definitions in Article 1 and
procedures outlined in Article 11, the “date of bringing back into use” of a frequency
assignment to a space station in the GSO is the actual date on which the frequency
assignment is resumed and the station begins transmitting again as per its notified
characteristics.
\end{vanillabox}
\end{subfigure}\hfill
\begin{subfigure}{0.485\linewidth}
\begin{ragbox}
\textbf{Answer:} B.\ The day on the commencement of 90-day period defined below. \Good

\medskip
\small \textbf{Explanation:} From the provided context:
\small
“The date of bringing back into use of a frequency assignment to a space station in the
geostationary-satellite orbit shall be the commencement of the 90-day period defined below.”
This language (e.g., in sections 20\textit{bis}, 24\textit{bis}, 3211.49.1) explicitly defines the
start of the 90-day period as the official date of bringing back into use.
\end{ragbox}
\end{subfigure}

\caption{Multiple-choice question with vanilla vs.\ RAG answers (bringing back into use date).}
\label{fig:bring-back-into-use}
\end{figure}

\begin{figure}[htbp]
\centering

% --- Row 1: full-width question + options ---
\begin{subfigure}{\linewidth}
\begin{qbox}
\textbf{Question:} What is the frequency band that is identified for use by high-altitude
platform stations (HAPS) in accordance with the provisions of Resolution 166 (Rev.WRC-23)?

\medskip
\textbf{Options}
\begin{enumerate}[label=\textbf{\Alph*.}, leftmargin=*, itemsep=2pt, topsep=2pt]
  \item 25.25--27.5 GHz.
  \item 27.0--27.5 GHz
  \item 27.5--29.0 GHZ
  \item 29.5--31.75 Hz
\end{enumerate}
\end{qbox}
\end{subfigure}

\vspace{0.5em}

% --- Row 2: two columns for answers ---
\begin{subfigure}{0.485\linewidth}
\begin{vanillabox}
\textbf{Answer:} A.\ 25.25--27.5 GHz  \Bad

\medskip
\small \textbf{Explanation:}
According to Resolution 166 (Rev.WRC-23) of the ITU, the frequency band
25.25--27.5\,GHz is identified for use by High-Altitude Platform Stations (HAPS).
This band is part of the fixed service spectrum and is suitable for broadband
connectivity via HAPS systems, which operate in the stratosphere to provide
telecommunication services.
\end{vanillabox}
\end{subfigure}\hfill
\begin{subfigure}{0.485\linewidth}
\begin{ragbox}
\textbf{Answer:} B.\ 27.0--27.5 GHz  \Good

\medskip
\small \textbf{Explanation:}
According to the provided context, Resolution 166 (Rev.WRC-23) identifies the
following bands for HAPS:
\begin{itemize}[leftmargin=*]
  \item 25.25--27.0\,GHz for ground-to-HAPS direction.
  \item 27.0--27.5\,GHz for HAPS-to-ground direction.
\end{itemize}
This makes Option B (27.0–27.5 GHz) the correct answer.
\end{ragbox}
\end{subfigure}

\caption{Multiple-choice question with vanilla vs.\ RAG answers (HAPS identification band).}
\label{fig:haps-band}
\end{figure}

\end{document}